\documentclass[twocolumn,showpacs,showkeys,preprintnumbers,amsmath,amssymb,
prb]{revtex4}

\usepackage{graphicx}
\usepackage{dcolumn}
\usepackage{bm}

\begin{document}

\title{Dynamics in a cluster under the influence of intense femtosecond 
hard x-ray pulses}

\author{Zolt\'{a}n Jurek}
\author{Gyula Faigel}
\author{Mikl\'{o}s Tegze}
\affiliation{Research Institute for Solid State Physics and Optics\\
	     H 1525 Budapest, POB 49, Hungary}

\date{January 27, 2003}

\begin{abstract}
In this paper we examine the behavior of small cluster of atoms in a
short (10--50 fs) very intense hard x-ray (10~keV) pulse. We use
numerical modeling based on the non-relativistic classical equation of
motion. Quantum processes are taken into account by the respective
cross sections. We show that there is a Coulomb explosion, which has a
different dynamics than one finds in classical laser driven cluster
explosions. We discuss the consequences of our results to single
molecule imaging by the free electron laser pulses.
\end{abstract}



\maketitle

\section{\label{sec:intro}Introduction}

Recently, there is a growing interest in the interaction of high
intensity electromagnetic field with solids and small clusters of
atoms. This interest is driven by two sources: i. by the availability
of short-pulse high-power laser sources in the few hundred nm
wavelength range, and ii. by the fact that the building of linac-based
free-electron laser type hard x-ray sources became a
reality.\cite{LCLS,JArthur} In the case of the long wavelength laser
radiation the interest shifted from the bulk-field to the
cluster-field interaction. The cause of this is that the behavior of
clusters under the influence of intense laser field shows several
peculiarities. First, there is a photo induced Coulomb
explosion.\cite{1Coulomb} Second, the interaction is much more
energetic than that of isolated atoms.
\cite{DitmNature386,ShaoPRL} Third, highly ionized states of the
atoms appear.\cite{DitmNature386} Fourth, lately it was shown that
even laser driven nuclear fusion could take place in small clusters of
deuterium atoms.\cite{DitmNature398} This could eventually lead to
the development of tabletop neutron sources. 

On the other hand, short-pulsed intense hard x-ray sources also
promise unique applications: the high energy of these photons allows
time dependent spectroscopic investigations of deep atomic levels, the
short wavelength makes possible structural investigations with atomic
resolution on a time scale of 100~fs. So we can follow chemical
reactions and biological processes in time. One can even think of
imaging individual molecules, viruses or clusters of atoms and
molecules using the very intense and short pulses of these x-ray
sources.\cite{Neutze} Further, we could study exotic states of matter
such as warm dense matter, etc.\cite{warmdens1,warmdens2} The
treatment of the high-energy case requires a very different approach
than the low photon energies. The reason is twofold: practical and
theoretical. From the practical point of view the production of an
intense hard x-ray photon beam requires many km long facility, which
costs hundreds of millions of dollars, while a high power infrared
laser source can be realized in normal laboratory environment for less
then a million dollar. This difference results in a very different
research strategy. While in the long wavelength case theory and
experiments develop parallel, in the x-ray case no experiment can be
done presently. However, there is a strong need for model
calculations, which can predict the behavior of different forms of
matter under the influence of intense x-ray beam. This information is
necessary, because planning these large machines and working out the
scientific case, one has to know how the optical elements and the
sample will behave in the beam. Based on this we can plan future
experiments and determine what kind of information can be gained from
them.

This leads to the theoretical side. It is clear that in the x-ray
case, the high energy of a single photon allows direct interaction
with core electrons, so the ionization mechanism significantly differs
from that of the low energy laser photons. This requires a different
theoretical approach. We expect that quantum mechanics and quantum
electrodynamics should be more often invoked than in the low energy
case. 

At the same time we know that the exact quantum treatment of a
thousand particle system in intense electromagnetic field is out of
the reach of present day computer capabilities. Therefore one has to
find a border where quantum and classical description meet, meanwhile
the behavior of the model system is not distorted significantly. The
first steps in this direction have been done. There have been model
calculations, which statistically describe a system after a single
primary ionization event.\cite{Beata1,Beata2}

In this paper we describe a model calculation for the dynamics of
atoms, ions, and electrons in a cluster, during an intense hard x-ray
pulse. In this case many consecutive ionization (energy deposition)
events in the system drive the cluster to highly ionized states,
leading to Coulomb explosion. In the calculations we work with
classical particles solving the classical equation of motion. The
quantum mechanics and the quantum electrodynamics are included through
cross sections. In practice it means that the various events are taken
into account by different probabilities. The motivation of these
calculations comes from two sources: in one hand we would like to see
the difference in the Coulomb explosions caused by low and high-energy
photons. Secondly, we would like to examine, how realistic the single
particle imaging by high intensity fs x-ray pulses is. In this article
we concentrate on the Coulomb explosion and give only a brief comment
on single particle imaging. The reason for this is that the imaging
problem is very complex, it requires not only the knowledge of the
dynamics of the cluster, but also the calculation of the intensity
distribution of elastic and inelastic x-ray scattering and the
reconstruction of atomic order from the scattering pattern. These
questions will be addressed in forthcoming papers.

\section{\label{sec:model}Model}

In order to do a realistic modeling we have to use input parameters
typical for the future linac based x-ray sources. Therefore, we give
the relevant characteristics of these sources below: the pulse shape
is gaussian with full widths at half maximum $FWHM=10$ and $50$~fs;
the number of photons/pulse is $N_{ph}=5*10^{12}$ ; the diameter of
the probe beam at the sample is $d=100$~nm (by focusing); the energy
of the beam is $E=10$~keV and it is linearly polarized. Before we
continue with the description of our model, we would like to introduce
a terminology: we will call a single x-ray pulse and all the events in
the cluster from the start of the pulse to the end of it an
\textquotedblleft experiment\textquotedblright.

Since our aim is the determination of the charge distribution in space
and time, we follow every individual particle (atoms, ions, and
electrons). This means the numerical solution of the classical
equation of motion for all particles. Quantum mechanics is included
via cross sections, and taking into account discrete atomic energy
levels in the ionization process. Here we would like to call the
attention to a difference between calculations in the low and
high-energy case. In the low energy case the laser field acts on the
particles in two ways: it can strip weakly bound electrons by
multiphoton process or optical tunneling, and it accelerates the ions
and electrons as a classical field. The charged particles move large
distances (compared to the cluster size) during half period of the
field, which can be taken as homogeneous within the cluster, since the
wavelength is much larger than the cluster size. In the x-ray case,
the field is changing very fast both in time and space, so that there
is no time for a particle to gain appreciable velocity and to move
large distance\cite{osccalc} during half period of the incident
beam. Therefore the x-ray field as a classical field can be neglected
in the equation of motions. So the most important interaction, which
alters the motion of charged particles, is the Coulomb
interaction. This is taken into account in our calculations and the
Coulomb interaction is not cut at any distance. However, close to the
nuclei the Coulomb potential is regularized for two reasons: we know
that in the vicinity of the nuclei the atomic electrons modify the
pure Coulomb potential. Secondly, in the classical picture an electron
could go very close to the positive nucleus and in this case the
potential diverges to infinity, which cannot be handled
numerically. Therefore in practice we use the following formula for
the Coulomb interaction: $U(r)=q/\sqrt{r^2+r^2_0}$, where $r_0$ was
chosen in a way, not to violate the energy conservation within the
numerical error.\cite{r0value}

The next approximation, which we have to mention, is the
non-relativistic approach. This is justified by the low maximum
velocity of electrons, which can be estimated from the incident photon
energy and the binding energy of the electrons. Taking the parameters
of the incident beam, the upper limit for electron energy is
10~keV. This corresponds to a velocity of about $1/5^{th}$ of the
velocity of light. Therefore the non-relativistic treatment is
justified. 

At last we would like to specify the cross sections. Analyzing the
possible scattering processes we arrive at two types of cross
sections: photon-particle and particle-particle. In the former we
include photon-electron, photon-atom, and photon-ion cross
sections. Photons with free electrons interact via Compton
scattering. The differential and total cross sections for this process
are given in quantum electrodynamics handbooks.\cite{Heitler} Using
the total Compton cross-section and the parameters of our experiment
we can estimate the number of Compton scattered photons during the
full length of the x-ray pulse. We get about 200 Compton scattering
events in a 1500 atom system. This low number means that Compton
scattering does not alter the time evolution of the charge
distribution at a detectable level. Therefore we neglect it in the
calculations.  In the case of strongly bound electrons the dominating
process is the photo effect. This is true for atoms and also for ions
provided that they have electrons left on deep core levels. Photo
effect cross section data for ions were extrapolated from the atomic
values.\cite{Heitler} Two approximations were used: first, we
neglected the change of the wave function of the atomic electrons on
removing electrons from the atom. Second, the probability of the photo
effect was normalized to one electron (at a given state), and
depending on the ionization state it was scaled by the number of
electrons actually present on the ion at the state under
consideration. The last possibility for the photon-atom interaction is
the fluorescent process. In our case (low Z sample) the probability of
the fluorescent decay is low compared to the Auger
process.\cite{Azaroff}

Considering the particle-particle type interactions, we can list
atom-atom, atom-ion, ion-ion, electron-atom, electron-ion, and
electron-electron interactions. We do not use explicitly cross
sections for atom-atom and atom-ion collisions. The reason is that
these collisions come into play only at the very beginning of the
x-ray pulse, since atoms are very rapidly ionized at the rising edge
of the pulse and between ions the Coulomb interaction dominates. For
those few atoms, which are not ionized the van der Waals type
interaction which we use to mimic chemical bonding describes well
enough the atom-atom and atom-ion collisions. 

Ion-ion and electron-electron interactions are taken into account
directly by the Coulomb interaction. Although this way quantum effects
(like exchange interaction) are neglected, we expect that at the given
experimental conditions (energy, density etc.), their contribution is
minor. 

The remaining two interactions, electron-atom and electron-ion are the
most important ones. We can distinguish three types: the Auger
process, the elastic scattering of electrons, and the secondary (often
called impact) ionization by electrons. In the Auger process an
electron from a higher level drops into a K hole (created previously
in the photo effect), meanwhile an other electron from the higher
level is emitted taking the excess energy. The lifetime for this
process was taken from Ref.~\onlinecite{KrauseAuger}. Probabilities
were scaled similarly to the photo effect, using the values given for
the basic Auger process.

The next interaction is the elastic scattering of electrons on atoms
and ions. This does not play an important role in the time evolution
of charge distribution. The reason is that this process does not
change the number of charges, only the direction of their velocity. We
checked the validity of the above statement by carrying out
calculations with different elastic cross sections. We tried both
isotropic and non-isotropic cross sections\cite{LandauIII} and we did
calculations without electron-atom elastic scattering. We found that
there was no significant difference in the time dependence of the
charge distributions among the three types of calculations. Therefore
we switched off atom-electron elastic scattering. We have to mention,
that part of the elastic scattering, namely the electron-ion
interaction is taken into account anyway by the Coulomb interaction,
which is present for all charged particles in the system.

\begin{figure}
\includegraphics[scale=1]{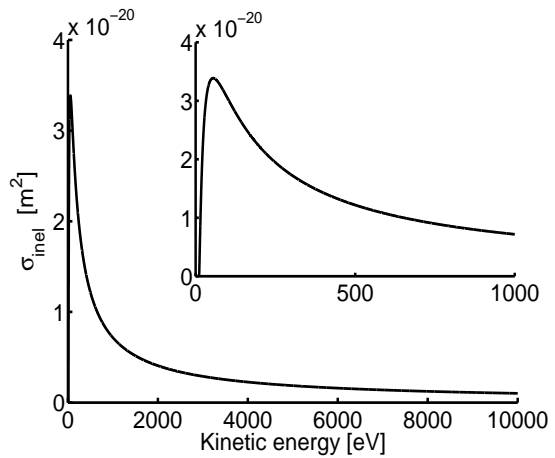}
\caption{\label{fig:inelcrossect}Cross section of electron impact ionization
vs. incoming electron energy for neutral carbon atoms. The inset shows
the cross section enlarged in the low energy regime. Note that for
ions the cross section is different according to
Eq.~(\ref{eq:siginel}). The mean free path dynamically changes because
of the inhomogen time-dependent atom-density.}
\end{figure}

In the secondary ionization an electron with high enough energy
interacts with an atom or ion kicking out an electron from a bound
state, meanwhile its kinetic energy decreases (so in this process the
number of charges changes). It is clear from the literature that with
the decrease of the energy the cross section of this process
increases. However, at very low energies (below 80~eV) this tendency
changes, and the incident electron energy dependence of the cross
section shows a rapid decrease (Fig.~\ref{fig:inelcrossect}).  All in
all in the lowest 80~eV region the description of the cross section of
the secondary ionization by electrons is problematic. Beside the
atomic properties the details depend strongly on the chemical bonding
and geometrical arrangement of atoms. Since in this region the cross
section values derived from different theoretical
approaches\cite{Tanuma1,Tanuma2,Ashley1,Ashley2} differ significantly
and experimental data are scarce we tried two approaches with 50 and
80~eV turning points for a 1500 atom cluster and 50~fs pulse width. We
found small differences in the cluster dynamics. These appeared close
to the end of the x-ray pulse. The changes was so small that they do
not effect the conclusions we draw from the calculations. Therefore in
all the other calculations we used the 50~eV turning point curve for
the cross section of the secondary electron ionization.

In what follows we describe the mechanism of the modeling. The
calculation proceeds via time steps. The typical value of a step is
$10^{-3}$ fs. In every time steps two actions are done:

~~~~~~~~~i.~Monte Carlo (M.C.)

~~~~~~~~ii.~Solving the equation of motion

In the M.C. sub-step we examine for all particles if any process
(taken into account by cross sections) happened or not. Therefore the
probabilities of the photo effect and Auger process are normalized to
one time step. For every atom and ion random numbers are generated to
decide if these processes had taken place. The secondary ionization is
calculated in a different way. First, the near neighbor atoms and the
relative atom-electron velocities are determined for every
electron. We attach to the electron a circular plate perpendicular to
its velocity. The area of this plate is equal to the cross
section. This plate moves together with the electron sweeping a
cylindrical volume. If in this volume there is an atom having bound
electrons with low enough energy to kick out, a secondary ionization
does happen.

At this point it is appropriate to discuss a delicate question: how we
create a new particle in our classical model. This is the point where
the classical and quantum picture have to be smoothly joined. It is
clear from the above description of the model that electrons appear in
the case of photo effect, Auger process, and secondary ionization. Let
us start with the simplest case of the photo effect. This seems
straightforward, since a photon comes in, and it looses all of its
energy, which is given to one atomic electron. This simple picture is
true if we look at the initial state as an atom at the origin and a
photon at infinite and the final state as an ion in the origin and a
kicked out electron at the infinite. However, in our calculations we
follow the path of every particle, and we know that the stripped
electron continuously moves out from the atom under consideration. On
its path it interacts with other atoms, ions, and electrons present in
the sample. Therefore we have to place the electron somewhere close to
its parent atom and not at infinite. This raises two questions: where
and with what velocity to put this electron. We use different values
for K and L electrons. These values can be directly calculated using
two plausibile assumptions: i. the distance and the magnitude of the
velocity are fixed for a given shell independently of the ionization
state of the atom; ii. conservation of energy is satisfied. This way
we arrive to distances about the Bohr radius, which is a natural
border of the atom in the classical picture. The velocity cannot be
simply $v=\sqrt{2E_{photon}/m_{el} -2E_{binding}/m_{el}}$ because the
electron should have this velocity at infinite distance from the
ion. If we put the electron close to its parent atom, which is now a
positive ion, the electron would slow down going to
infinity. Therefore we have to give the electron a larger velocity to
compensate this slowing down. If we want to be more precise, we have
to take into account the field caused by all other charged particles
(though in practice the leading term is the Coulomb potential of the
closest ion). The last problem is the direction of the electron's
velocity. The angular dependence of the cross section of the photo
effect is given in handbooks\cite{Heitler} for a linearly polarized
incident photon. Therefore we use random directions with a
distribution corresponding to the theoretical cross section. The
direction of the velocity fixes the position of the exact placement,
since we put the electron in a way that it moves out radially from the
atom. At last we have to mention that the energy and momentum
conservations are satisfied in this process, so the ion takes recoil
energy. For placing the electron in the Auger process, a similar
mechanism is used. However, in the case of secondary ionization there
is a significant difference. The cause is that here we have a three
body problem. In this case we have an electron and an atom (or ion) in
the initial state and two electrons and an ion in the final
state. Placing the electrons with the proper velocities is not
straightforward. The first problem is that there is no reliable data
for the angular dependence of the cross section of inelastic electron
scattering. The simplest assumption is an isotropic emission of the
secondary electrons and we use this in the calculations. To see the
effect of non-isotropic emission we did model calculations with an
angular distribution derived from quantum mechanical
calculations.\cite{LandauIII} There was a small change in the time
evolution of the system.  So our assumption of isotropic emission was
justified. For the energy dependence of the cross section we use a
parameterized formula\cite{YKKIM}:
\begin{eqnarray}
\sigma_{inel}=\frac{S}{t+\left(u+1\right)/n}
\left[\frac{Q \ln t}{2} (1-\frac{1}{t^2})\right.+\nonumber\\
\left.\left(2-Q\right)\left(1-\frac{1}{t}-\frac{\ln t}{t+1}\right)%
\right]%
\label{eq:siginel}
\end{eqnarray}
where T, U, B, and N are the energy of the incident electron, the
orbital kinetic energy, the binding energy, and the electron
occupation number respectively, $t=T/B$, $u=U/B$,
$S=4{\pi}a_0^2N(R/B)$, $a_0=0.529*10^{18}~\textrm{\AA}$, $R =
13.6057~eV$, the dipole constant $Q$ is approximated with 1, and n is
a value near 1 used for ions.\cite{YKKIM}

The above parameters are based on experimental data.\cite{YKKIM}
Further, we correct this velocity to get the local velocity in a
similar way as it was done for the Auger and photoelectrons. Now we
should fix the velocity and position of the kicked out
electron. However, this cannot be done simply, because we have a three
body problem, and the energy and momentum conservation do not
determine unambiguously the velocity of the ion and the primary
electron after scattering. Therefore we assume, that the scattering of
the incident electron is in the plane determined by the velocity of
the primary electron before scattering and the vector pointing from
the nucleus to the primary electron.

At this point the M.C. sub-step is finished. The next sub-step is
solving the equation of motion. First we calculate the resultant force
for every particle. Starting from the forces, the new velocities and
positions are calculated using the fourth order Runge-Kutta method.

At the end of this section we describe the clusters investigated in
this study. As model systems we chose monatomic all carbon
clusters. The atoms were held together by simple central forces
only. We used the following potential function:
\begin{eqnarray}
V(r)=V_C\left[\left(\frac{\sigma}{r}\right)^6-1\right]%
\left(\frac{\sigma}{r}\right)^6%
\label{CCbond}
\end{eqnarray}
where the values of $V_C$ and $\sigma$ parameters are chosen to have
the minimum of the potential at $r=1.5~\textrm{\AA}$ with the depth of
3.5~eV.  Starting atomic positions were simple cubic or face-centered
cubic ordered, or the above but with randomized positions about the
lattice sites (we used $max(|\Delta a|/a)= 0.05$ ,where $\Delta a$ is
the deviation from the ideal lattice site, while \textquotedblleft
$a$\textquotedblright\ is the lattice spacing). We found that the
actual starting atomic configuration hardly alters the explosion
dynamics, as far as the first neighbor distance is kept the
same. Therefore in what follows we show the results of calculations on
clusters with the simple cubic atomic order.

\section{\label{sec:results}Results} 

Before we show statistics, distributions, energy spectra etc. we would
like to outline what type of information we seek and how we estimate
the precision of our predictions.\cite{notestartconf} Our aim is to
map the characteristic behavior of clusters as a function of the size
of the cluster (number of particles) and the length of the x-ray
pulse. In size we covered the range from 50--1500 atoms/cluster. The
pulse widths are $FWHM=10$ and $50$~fs. The most important features
what we are interested in are: the total number of stripped electrons,
the spatial and energy distributions of atoms, ions, and electrons,
and the number of stripped electrons in the beam.

To see the effect and the importance of different interactions
governing the time evolution of the system, we did three types of
calculations. In these we turned on different interactions step by
step. In the first one (referring later as model I.) there are photo
effect, Auger process, and Coulomb interaction between ions. However,
the photo and Auger electrons leave the system without any interaction
(we repeated the calculation of Neutze et al.\cite{Neutze} for our
model system). It is clear that in these calculations we make two
errors: first we underestimate the number of stripped electrons, since
the photo and Auger electrons do not kick out further electrons from
the atoms and ions. Secondly the rate at which the charge state of the
cluster as a whole increases, is overestimated. This comes from the
fact, that the positive ions attract electrons, so that the slower
Auger electrons are unable to escape. At later times (i.e. for large
charge state of the cluster, $Q>10^4e$) even the faster photoelectrons
are significantly slowed down decreasing the temporary net charge of
the cluster.

The underestimate in the number of stripped electrons means that the
radiation damage is larger in reality than in the calculation of
Neutze et al.\cite{Neutze} However, this damage means a change in the
charge state of the atoms and it does not necessary followed by a
change in the position of atoms. Actually we expect slower increase of
the positional disorder than predicted in Ref.~\onlinecite{Neutze}
because of the overestimated total charge. In a real system a slower
increase of the charge would lead to a milder Coulomb explosion.

In the second type of calculations (model II) we have the same
interactions as in the first one, however, not only the ions but also
the photo and Auger electrons interact by the Coulomb
interaction. With this modification we correct for the above-mentioned
overestimate. However, the total number of ionizations, which is
important from the point of view of plasma dynamics, stays much behind
the reality.

Therefore, in the third type of calculation (model III) we introduced
inelastic electron-atom and electron-ion scattering in addition to the
effects taken into account earlier. That results in the appearance of
secondary electrons. However, in this case the number of stripped
electrons is overestimated. This comes from the fact that in the
classical equation of motion, atomic and ionic orbits are not
quantized. Therefore a classical electron (stripped electron in our
calculations) can drop very deep into the potential well of an ion,
meanwhile the secondary electron takes the excess energy in the form
of kinetic energy. This results in more stripped electrons, as we
would have in reality. In order to compensate for this effect we did
not allow negative binding energies between the two electrons
participating in the process and their nearest neighbor ion. However,
even with this adjustment a slight overestimate is expected for the
number of stripped electrons, because electronic relaxation,
recombination is not taken into account. They were left out because
according to experiments\cite{Sultana,Safronova} and theoretical
estimates\cite{LandauIII}, these processes have very small
probabilities on the time scale we are interested in.

\begin{figure*}
\includegraphics[scale=1]{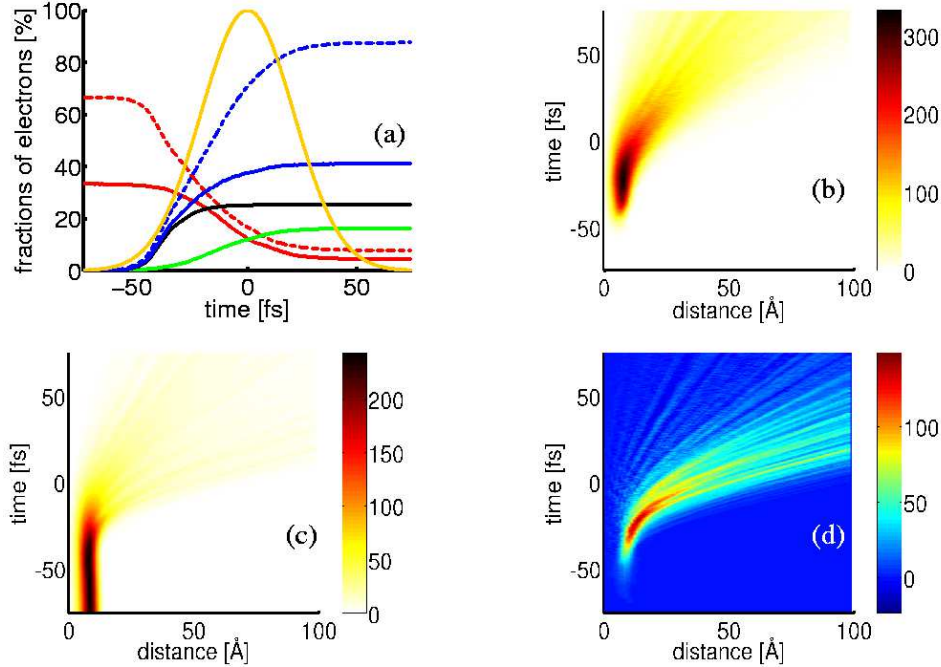}
\caption{\label{fig:spacetime}General properties of an exploding atomcluster 
in an x-ray pulse. The figures are for a 1500-carbon atom cluster in a
50~fs pulse. (a): Number of different types of electrons vs. time
(curves: solid red = K, dashed red = L, solid blue = stripped in the
beam, dashed blue = all stripped, green = Auger, black = secondary
electrons, and yellow = intensity of the beam). On (b), (c), and (d)
the radial distribution (number of particles in a 1 \AA\ thick
spherical shell with radius r) of electrons, atoms/ions, and the total
charge are shown respectively. Note that the middle of the pulse is at
t = 0~fs. We use this convention for the time in all figures.}
\end{figure*}

\begin{figure*}
\includegraphics[scale=1]{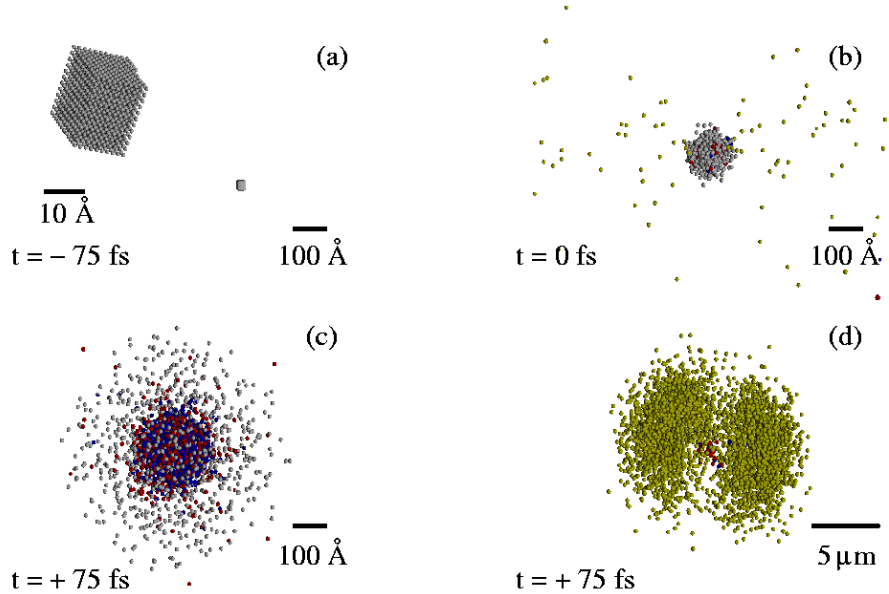}
\caption{\label{fig:realspace} An exploding 1500-atom cluster in real space
 at different times. The spheres with color gray, light green, red,
 and blue symbolize the atoms , photo- , auger-, and
 secondary-electrons, respectively. At the end of the pulse the radius
 of the cluster is about 15 times larger than it was originally (not
 including the photoelectrons). For better visualisation we show the
 initial configuration of the cluster enlarged on the upper left part
 of (a). We used opposite zooming on (d) in order to show the
 photoelectrons escaped far away from the cluster. Mostly Auger and
 secondary electrons concentrate at the center (c), whereas
 photoelectrons are leaving the system forming a butterfly-shaped
 cloud (d), reflecting the anisotropy of the photoeffect in the
 linearly polarized x-ray.}
\end{figure*}

The results we present are based on hundreds of calculations. Beside
doing calculations for the different models and various cluster sizes,
we also followed several independent explosions with the same
parameters except using different series of random numbers. This way
we could check the sensitivity of parameters to the stochastic nature
of the processes. We found that the statistical uncertainty was about
5\% for the 50-atom clusters and this fluctuation significantly
decreased for larger clusters. The various parameters shown in the
following figures are the result of averages of independent
explosions.It is clear that we cannot show all the curves and real
space distributions. Therefore, first we show typical results of one
model calculation (Fig.~\ref{fig:spacetime}, Fig.~\ref{fig:realspace})
and explain the main features of these figures. Then in the next part,
the different types of calculations will be compared and at the end
the dynamics of the Coulomb explosion will be given for various
cluster sizes and pulse lengths based on the most realistic model.

\begin{figure*}
\includegraphics[scale=1]{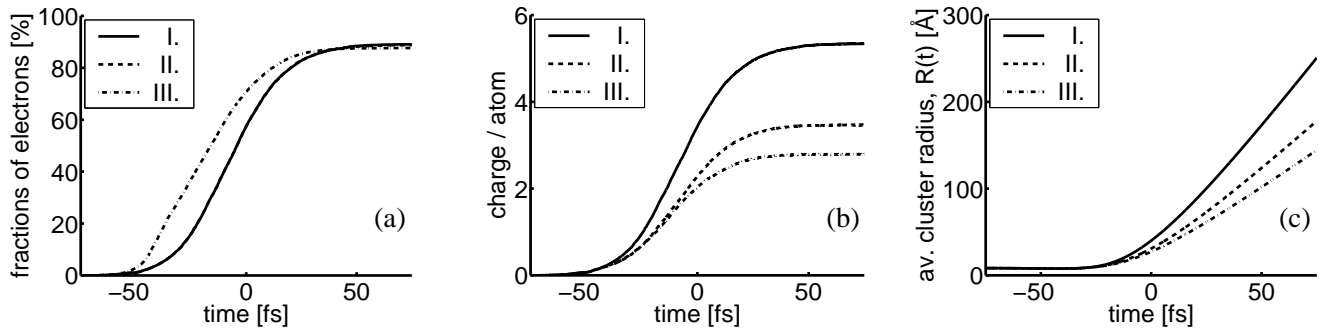}
\caption{\label{fig:diffmodels} The time evolution of the
number of stripped electrons, the charge of the cluster / atom, and
the average radius of the cluster are shown on (a), (b), and (c),
respectively, for an exploding 1500-atom cluster.  The three different
curves denote the three types of model calculations: (I) with photo
and Auger electron emission, without secondary ionizations, neglecting
the presence of the non-bounded electrons; (II) including the photo
and Auger electrons, but still excluding secondary ionizations; and
(III) including secondary electron emission in addition to the
interactions taken into account in I and II. Note that on (a) curve I
and II exactly coincide.}
\end{figure*}

\begin{figure*}
\includegraphics[scale=1]{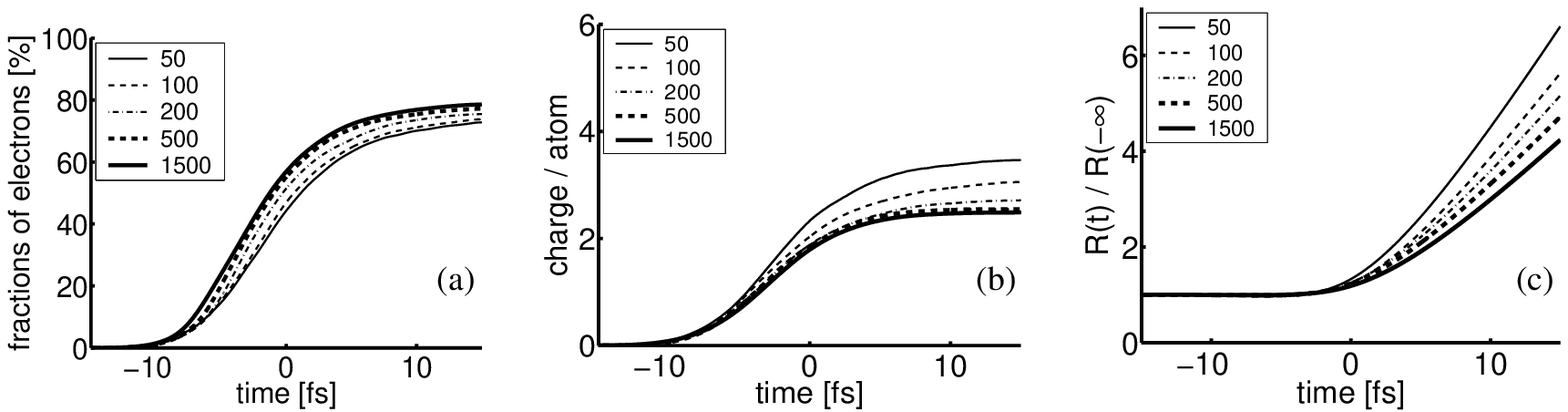}
\caption{\label{fig:params10} Time evolution of the number
of stripped electrons (a), the average total charge of the cluster per
atom (b), and the radius of the cluster (normalized by the initial
cluster radius) (c), in a 10~fs x-ray pulse. Various curve types
correspond to clusters containing different number of atoms.}
\end{figure*}

\begin{figure*}
\includegraphics[scale=1]{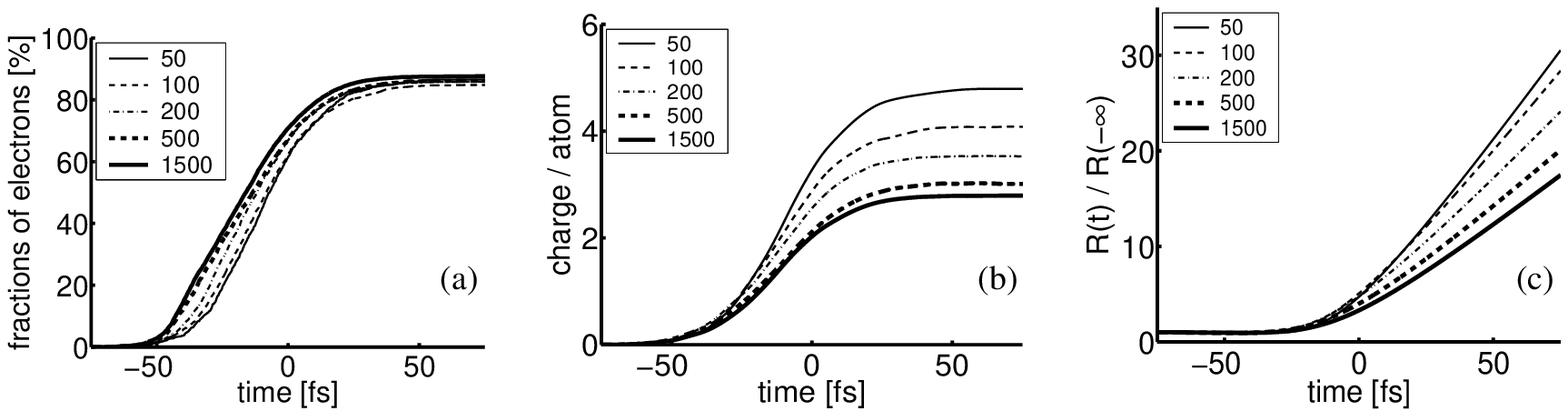}
\caption{\label{fig:params50} Time evolution of the number
of stripped electrons (a), the average total charge of the cluster per
atom (b), and the radius of the cluster (normalized by the initial
cluster radius) (c), in a 50~fs x-ray pulse. Various curve types
correspond to clusters containing different number of atoms.}
\end{figure*}

We have chosen for demonstration a 1500-carbon atom cluster. The atoms
are arranged in a simpe cubic lattice with lattice spacing of
$a=1.5\textrm{\AA}$ and the starting atomic positions randomized about
the regular lattice sites as described earlier. The pulse length is
50~fs and $5*10^{12}$ photons are in a pulse. According to the
categorization given earlier, the calculation is in the third category
(all interactions are included). On Fig.~\ref{fig:spacetime}(a) we
show the number of different type of particles as a function of
time. Since we do classical calculations, we can flag electrons by
their origin. That allows us to distinguish Auger, photo and secondary
electrons. The curve labeled stripped electrons in the beam needs
further explanation. It is calculated by counting the number of
electrons in the volume of a 1000~\AA\ diameter cylinder with its axis
coinciding with the x-ray beam. The significance of this curve is that
it allows to estimate the number of inelastically scattered photons of
the incident beam. All numbers are normalized by the total number of
electrons present in the sample at zero time. On
Fig.~\ref{fig:spacetime}(b), (c) and (d) we show the time evolution of
the spatial distribution of stripped electrons, ions and the total
charge respectively. The horizontal axis corresponds to the real space
distance from the center of mass. The vertical axis denotes the
time. The colors show the number of electrons in a spherical shell
with a width of 1~\AA\ radius corresponding to the values on the
horizontal axis. On Fig.~\ref{fig:realspace}. the cluster in real
space is shown as if it were photographed at different times. At this
point we do not want to analyze Fig.~\ref{fig:spacetime} and
~\ref{fig:realspace} in details but we point out a few features
characteristic for all calculations: (i) atoms lose a significant
number ($\sim70\%$) of their bound electrons within the first half of
the x-ray pulse, (ii) photoelectrons leave the cluster shortly after
their emission (but a significant number are in the beam), (iii) The
butterfly shaped spatial distribution of photoelectrons reflects the
direction of polarization of the incident beam. (iv) The cluster
looses its nuclear topology very early during the pulse (well before
half of the photons hit the sample).

After introducing the parameters, which characterize the explosion, we
compare finer details predicted by different models (the three
different types of calculations, introduced previously). We used the
same parameters of the cluster and the pulse as in the experiment
described in the previous paragraph. On
Fig.~\ref{fig:diffmodels}(a),(b), and (c) the number of stripped
electrons, the average charge/atom (the degree of ionization), and the
radius of the cluster
($R(t)=\sqrt{\frac{1}{N}\sum\limits_{i}^{N}{r_i(t)^2}}$, where $N$
denotes the number of atoms in the cluster and $r$ their distance from
the center of mass) are shown respectively. The figure clearly
reflects those features discussed earlier in this section: the first
type of calculation overestimates the increase of the charge
[Fig.~\ref{fig:diffmodels}(b)], and it underestimates the ionization
rate of atoms [Fig.~\ref{fig:diffmodels} (a)]. Including Coulomb
interaction for all charged particles (second type of calculation) we
correct somewhat for the overestimate. This is best seen on the charge
[Fig.~\ref{fig:diffmodels}(b)], but it also changes the dynamics of
the Coulomb explosion, see the time dependence of R(t)
[Fig.~\ref{fig:diffmodels}(c)]. We expect that the true behavior of
the cluster is closest to the third type of calculations in which all
interactions are taken into account. Therefore, in what follows we
show the results of the third type of calculations only.

First we analyze the dynamics of the Coulomb explosion as a function
of cluster size and pulse length. In Fig.~\ref{fig:params10} and
~\ref{fig:params50} we depicted the time dependence of the number of
stripped electrons, the average charge/ion, and the normalized cluster
size ($R_N(t)=R(t)/R(-\infty)$ ) for systems containing 50, 100, 200,
500, or 1500 particles. The pulse length was 10~fs and 50~fs
respectively.

Let us start with the discussion of Fig.~\ref{fig:params10}(a) and
~\ref{fig:params50}(a). There is a trend in the number of stripped
electrons with the cluster size: the larger the cluster the faster the
ionization. This can be explained by the secondary ionization, since
for small clusters there is a larger chance for Auger and
photoelectrons to leave the system without kicking out another
electron. For first sight Fig.~\ref{fig:params10}(b)
[~\ref{fig:params50}(b)] and ~\ref{fig:params10}(c)
[~\ref{fig:params50}(c)] contradict Fig.~\ref{fig:params10}(a)
[~\ref{fig:params50}(a)], since the normalized size of the clusters
and the average charge of the particles increase more slowly for
larger clusters. The explanation of this effect is that photo and
Auger electrons loose their energy on secondary ionization and their
remaining kinetic energy is not enough to escape from the Coulomb
attraction of the positive net charge of the cluster. These slow
electrons shield the ion-ion repulsive Coulomb interaction, leading to
slower Coulomb explosion of larger clusters.

\begin{figure}
\includegraphics[scale=1]{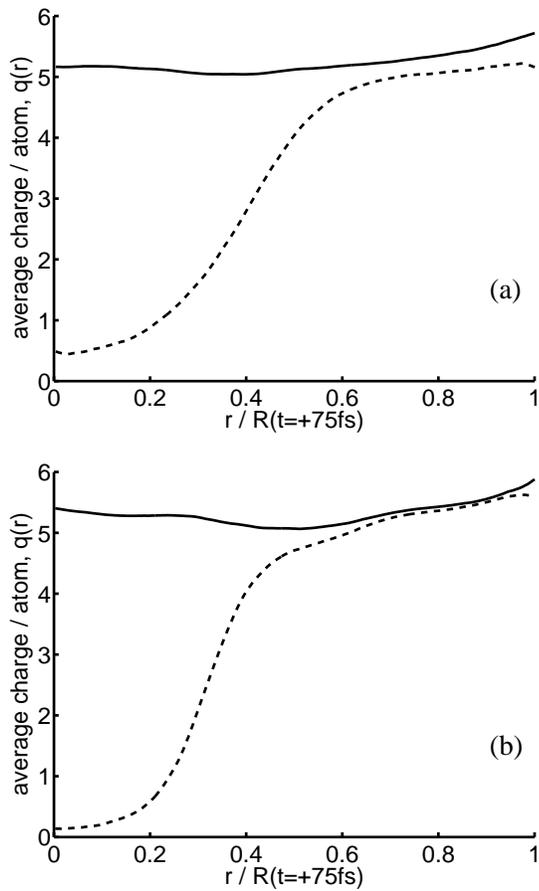}
\caption{\label{fig:chdistr} Spatial distribution of the average ionic
 charge per atom (solid line) and the clustercharge per atom (dashed
 line) in a cluster containing 500 (a) and 1500 (b) atoms, at the end
 of the pulse. Note that the distance from the center of mass is
 normalized for easy comparison.}
\end{figure}

\begin{figure}
\includegraphics[scale=1]{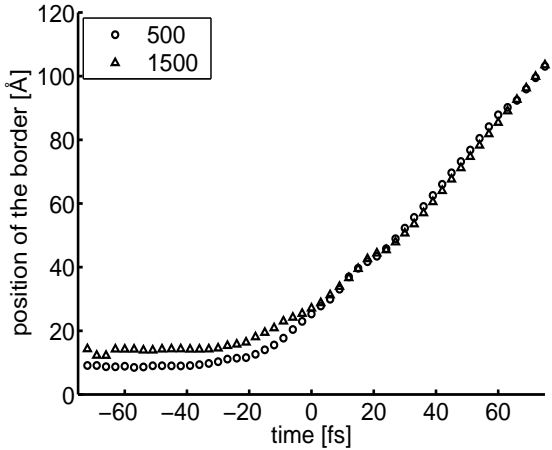}
\caption{\label{fig:border} Radius of the inner part of the cluster vs. 
time. }
\end{figure}

Beside the time dependence of the number of \textquotedblleft
free\textquotedblright\ charges, their spatial distributions are also
important characteristics of the explosion. In Fig.~\ref{fig:chdistr}
the radial charge distributions are shown for a 500 (a) and a 1500 (b)
atom cluster, at the end of a pulse with $FWHM=50$~fs. The behavior of
the systems are similar in the case of a $10$~fs pulse, therefore we
do not show that figure separately. Note that there is a step in the
charge distributions. We have an almost neutral plasma in the central
part of the cluster and a highly charged shell around it. This charge
distribution is formed from highly charged individual ions and
electrons. To illustrate this, we depicted the charge distribution of
ions independently. Note, that the degree of ionization is almost
constant everywhere, it does not follow the net charge
curve. Therefore the charge distribution is formed from highly charged
ions and electrons and it is not the result of a high concentration of
neutral atoms in the center part. In the following we discuss this
peculiar charge distribution in more details. We show that the two
spatially separated regions have different characteristics in the
energy domain and that the border of these regions moves out
continuously. We demonstrate this for large clusters, but one finds
similar behavior for smaller ones. We chose large systems because for
small clusters (50--100 atoms) the definition of the border between
the two regions is less precise.
\begin{figure*}
\includegraphics[scale=1]{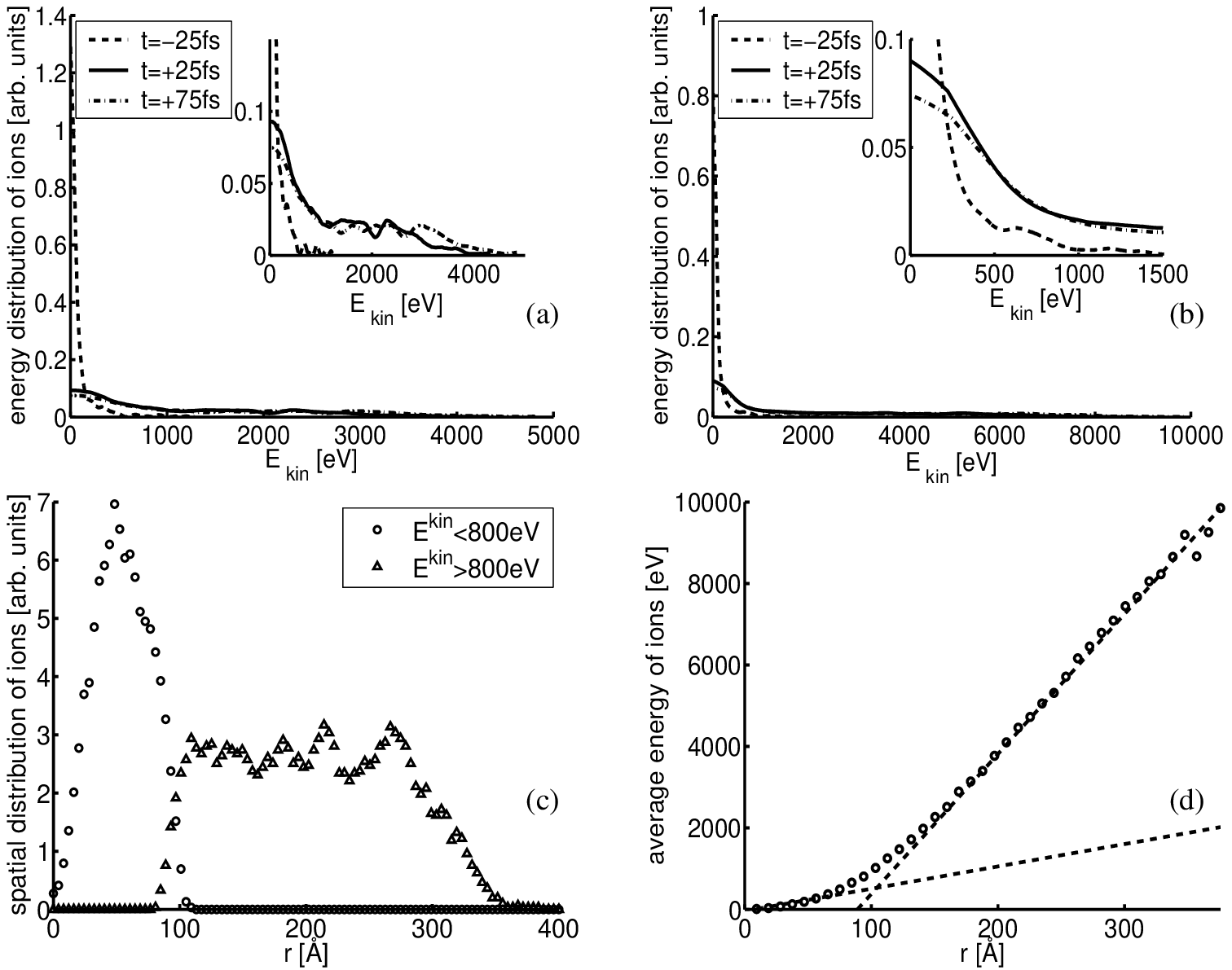}
\caption{\label{fig:Eatoms} Kinetic energy distributions of atoms at
different times for 500 (a) and 1500 (b) atom clusters with enlarged
energy regions in the insets. On (c) the radial distributions of
atoms in the peak and in the tail are plotted using open circles and
open triangles, respectively ($N=1500$, $t=+75~fs$,
$T_{pulse}=50~fs$). On part (d) the radial distribution of the average
kinetic energy is plotted. This curve can be described by two linear
functions shown as dashed lines.}
\end{figure*}
Fig.~\ref{fig:Eatoms}(a) and (b) show the energy distributions of ions
at various times during the pulse for a 500 and 1500 particle cluster,
respectively. There are two characteristic features: (i) The energy
scale is going up to the many keV range, and (ii) there are two
regions in energy, a low energy peak and a long high energy tail. We
defined a border between the two regions as the zero crossing of a
straight line fitted to the steepest part of the energy distribution
curve. Note that at the beginning of the pulse there is a time
interval (or a minimum number of incident photons), for which the two
regions cannot be defined, because there is not enough time and free
electrons and ions to form these regions. In this particular case this
interval extends to about $-15~fs$. We examined the characteristics of
these regions separately. First we show the spatial distribution of
the ions. In Fig.~\ref{fig:Eatoms}(c) a typical curve of the number of
ions as a function of the distance (r) from the center of mass is
shown. The number of ions was calculated by counting the ions in
1~\AA\ thick spherical shells with different radiuses (r). The spatial
distribution shows a similar shape as the energy distribution. It
seems that the resemblance is not accidental, the separation in energy
is connected to a separation in space: the low energy ions are close
to the center, in the first peak; up to 80~\AA. The ions in the high
energy tail are further away, in the tail of the spatial distribution
curve. This stronger correlation can be observed by plotting the
average energy of ions as a function of the distance from the center
of mass [Fig.~\ref{fig:Eatoms}(d)]. A monotonic increase of the energy
with the distance is observable. One can fit this curve by the sum of
two linear functions. Their crossing coincides with the border of the
almost neutral plasma and the highly charged outer shell, further
emphasizing the distinction between the two regions. We can get an
estimate for the speed of the Coulomb explosion from the motion of the
border. We plotted the position of the border as a function of time on
Fig.~\ref{fig:border}. According to this, the border moves out with
about 1.1~\AA/fs velocity, which corresponds to $\sim700~eV$ ion
energy.

\begin{figure*}
\includegraphics[scale=1]{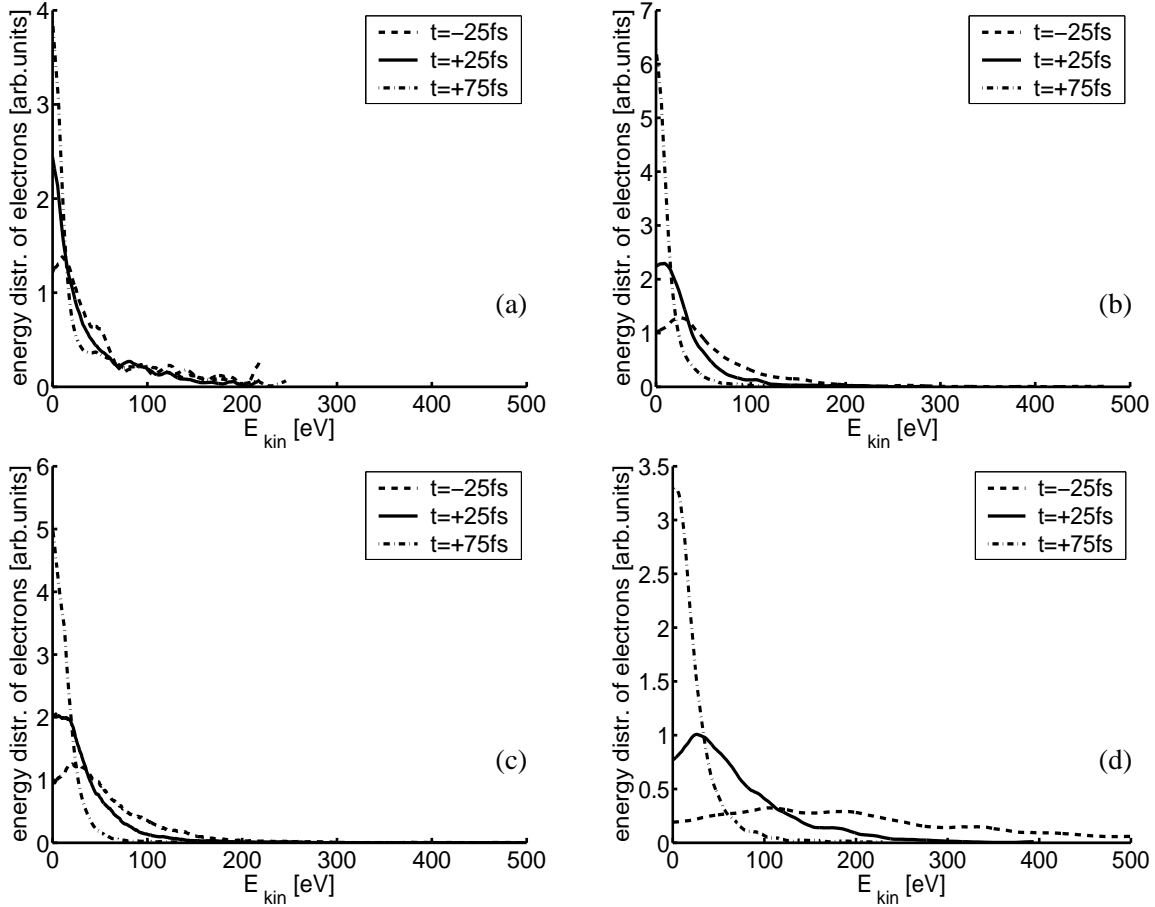}
\caption{\label{fig:Eelec} Kinetic energy distributions of electrons inside
 the cluster at various times based on (a) 100-atom, (b) 500-atom, and
 (c) 1500-atom calculations in a $FWHM=50~fs$ pulse. For comparison on
 Fig.~\ref{fig:Eelec}(d) we show the same distribution for a
 calculation excluding secondary ionizations ($N=1500$,
 $FWHM=50~fs$).}
\end{figure*}

So far we examined the spatial and energy distribution of ions. In the
next part we characterize the electrons. In
Fig.~\ref{fig:Eelec}(a)--(c) the energy distributions of electrons are
shown for various cluster sizes (100, 500, and 1500) and at different
times for pulse width of 50~fs. The combined effect of secondary
ionization and expansion of the cluster is observable. As we approach
the end of the pulse the energy distribution of electrons gets
narrower, more electrons \textquotedblleft condense\textquotedblright\
to small energies. This effect is more pronounced for larger systems,
in which the secondary ionization is more effective. In order to
estimate the effect of secondary ionization, we calculated the energy
distribution for models without the secondary ionization. The result
is shown in Fig.~\ref{fig:Eelec}(d) for a 50~fs pulse width and 1500
particle system. Comparing this distribution to that of
Fig.~\ref{fig:Eelec}(c) (the same system with secondary ionization),
differences can be observed. The first general impression is that much
less electrons are stripped. This is not surprising since we turned
off an ionization process. The second feature is that the
distributions get wider. This feature is more pronounced at the early
time of the pulse.

\begin{figure*}
\includegraphics[scale=1]{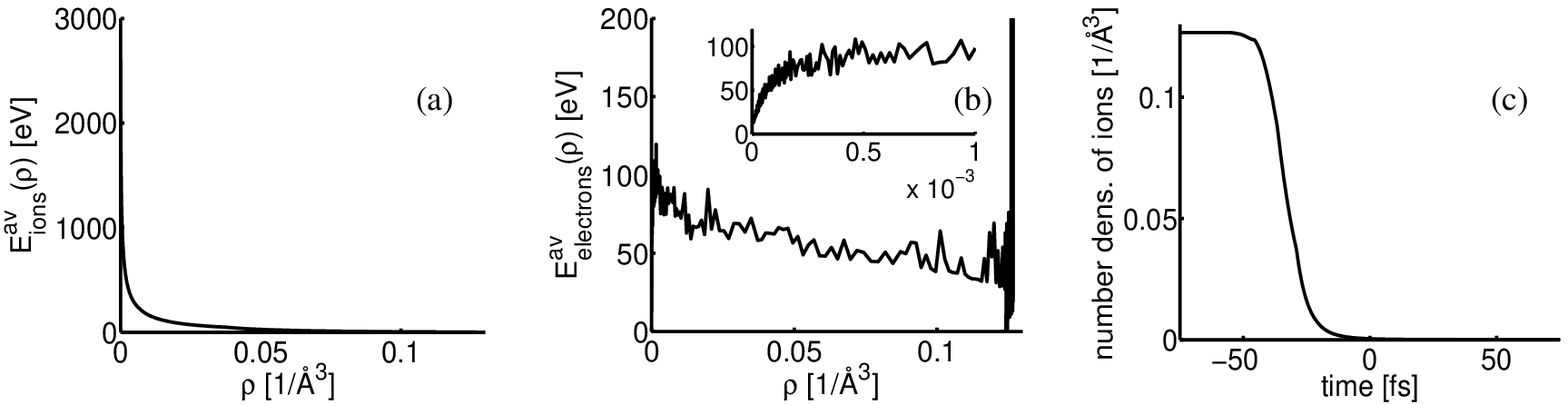}
\caption{\label{fig:density} Average kinetic energy of atoms (a) and 
electrons (b) inside the cluster vs. average atomic density
$\rho$. Part (c) shows the time dependence of the average atomic
density ($N=1500$, $FWHM=50~fs$).}
\end{figure*}

Beside the above features of the energy distributions we can try to
connect the energy to thermodynamic parameters. It is clear that the
system is small and far from equilibrium, so thermodynamic parameters
are hard to define. In spite of this, the knowledge of the relation
between density and kinetic energy (temperature) might give a clue to
the understanding of the governing processes. Therefore we depicted
the average kinetic energy of ions and electrons inside the cluster as
a function of the ion density, and the ion density as a function of
the time, on Fig.~\ref{fig:density}(a), (b), and (c). respectively.
The $E(\rho)$ function of ions is characteristic for an exploding
system, which we pump energy into. The energy has a rapid nonlinear
increase at low densities. The electrons behave very differently, they
are almost decoupled from the ions. Their energy is slightly, linearly
decreasing with the density. A deviation from this behavior can be
seen at very low and high densities. On the low density region there
is a drop [see Fig.~\ref{fig:density}(b)] and at high densities there
is a fluctuation in the energy. The drop is caused by the simultaneous
effect of two factors: the decrease of energy deposition by the
incident photon beam and the expansion of the cluster. The fluctuation
is a result of the low number of stripped electrons and the very high
energy of photoelectrons. High density appears at the early time of
the pulse. However, this time the number of stripped electrons is very
low, and mostly photoelectrons are present. If one or more of these
high energy ($\sim$10~keV) photoelectrons are in the cluster, where we
calculate density and energy distribution, we get a high
average. However, as the number of the secondary and Auger electrons
increase, the weight of photoelectrons becomes small. On Fig.
\ref{fig:density}(c) the time dependence of the ion density is shown. There
is a narrow time interval (centered about $-20$~fs), where a large
drop in the density takes place.

Before we finish the discussion of the results of our model
calculations, we compare our findings to calculations published
earlier. Unfortunately, there are not many works on this topic. The
closest to our modeling is the calculation of Neutze et
al.,\cite{Neutze} which we have mentioned earlier in the
paper. However, the aim of that work was not to give a detailed
analysis of the plasma and its time evolution but to get an impression
on the feasibility of single molecule imaging. Since in that paper
there was no detailed data on the spatial, time and energy
distributions of particles, we repeated that type of calculations with
the same parameters we used in our modeling (these calculations
correspond to the first type of modeling, according to our
categorization introduced earlier). The basic differences between our
modeling and the calculation performed in Ref.~\onlinecite{Neutze} is
that we follow all stripped electrons (we do not remove them from the
system), and all charged particles interact by Coulomb interaction. In
addition to this, the electrons are also inelastically scattered by
atoms and ions. These differences have two consequences: in one hand
the ionization of atoms are faster, but at the same time the Coulomb
explosion is moderated by the electron cloud formed by slow
electrons. This results in a different explosion dynamics, and spatial
charge and energy distributions. We find an almost neutral central
core expanding by 1.1~\AA/fs velocity, and a positively charged shell
formed by fast highly ionized ions about this core. In Neutze 's model
this type of charge distribution does not develop.

Recently, there has been another publication on cluster dynamics at
x-ray energies.\cite{Saalmann} However, in this case the cluster size
was small (13--55 Argon atoms) compared to ours (50--1500 Carbon
atoms), and the photon energy much lower (350~eV) than in our modeling
(10~keV). Therefore our findings differ from those of
Ref.~\onlinecite{Saalmann}. However, there are two similarities:
ionization starts from inner shells, and the incident beam does not
give appreciable velocity to charged particles, and therefore it
causes negligible spatial oscillation.

The third type of works, which we can compare our calculations to are
the works on the classical laser driven Coulomb explosion of small
clusters. There have been extensive theoretical and experimental
investigations in this
area.\cite{Laser1,Laser2,Laser3,Laser4,LaserPlus} Comparing results we
find substantial differences in every respects: spatial charge
distribution, energy distribution and in time dependence. This is not
surprising, since the underlying processes are different. The
difference originates from the very long wavelength (hundreds of nm-s)
of the incident beam compared to our case (0.1~nm). The consequences
of the large wavelength are: at a given time the full cluster sees the
same field; the direction of this field changes slowly compared to the
time an electron moves about the cluster size; single photon energy is
not enough to strip bound electrons. Since in the x-ray case we are in
the opposite limits, the behavior of the cluster in the short x-ray
and laser pulse is very different. This difference already manifests
in the ionization process. While in the classical laser case the
ionization proceeds by multiphoton ionization from outer shells and it
is followed by impact ionization via the stripped fast electrons
accelerated by the field. In the x-ray case the field of the incident
beam does not play such an important role, since it does not
accelerate stripped electrons to high velocities. The cause of this is
that the field direction is changing so fast that charged particles
can not gain appreciable velocity in this period. Therefore Auger and
secondary electrons stay in the cluster, and as the cluster of ions
expands the electrons condense to low energies. These electrons slow
down the Coulomb explosion, especially in the inner core of the ion
cluster. This very different explosion dynamics also reflects in the
energy distribution of ions. In the x-ray case the typical ion
energies are much lower. They are in the 10~keV range as compared to
the hundreds of keV found in classical laser driven Coulomb
explosions.

\begin{figure}
\includegraphics[scale=1]{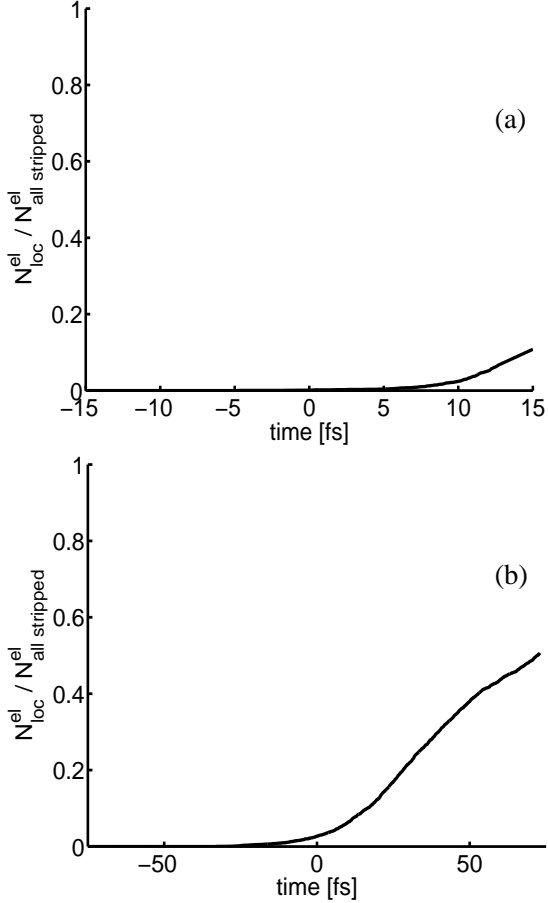}
\caption{\label{fig:e_loc} Fraction of electrons, which are classically
 localized for a 1500 atom cluster in a 10~fs (a) and in a 50~fs (b)
 pulse are shown as a function of time.\cite{locdef}}
\end{figure}
\begin{figure}
\includegraphics[scale=1]{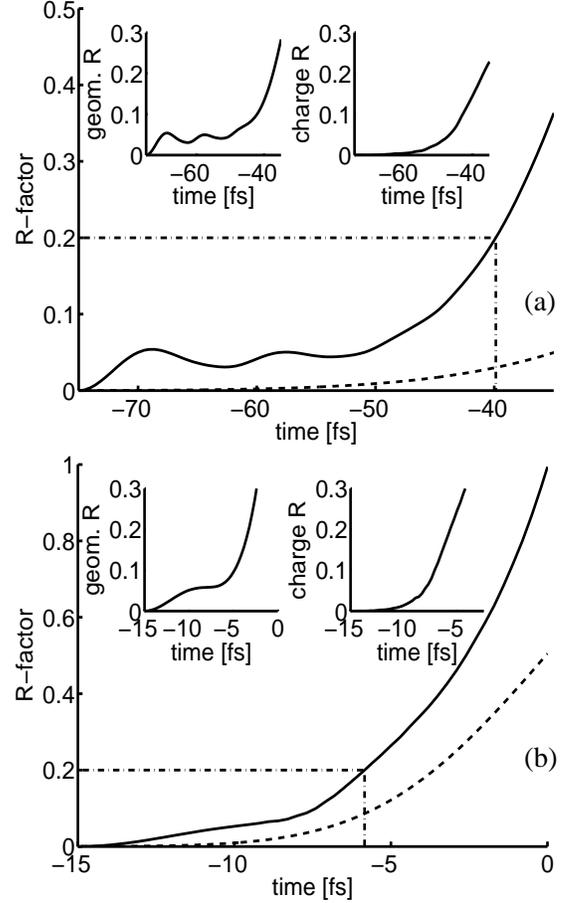}
\caption{\label{fig:Rfact} R-factors calculated from the geometrical 
distortion (geom.R) and from the changing of the average number of
atomically bound electrons (charge R) for a 1500 atom cluster in a
50~fs (a) and in a 10~fs (b) pulse. The insets show the geom R and
charge R separately. The dashed line shows the integral number of
photons.}
\end{figure}

After finishing the characterization of the Coulomb explosion, a few
words are appropriate about the time scale in which the validity of
our model calculation is justified. A limit is given by the
recombination processes, which we neglected. Recombination plays an
important role if a large ratio of the electrons localize about ions
with small energy. To see this, we counted how many electrons stay at
an ion for long periods.\cite{locdef} This is shown on
Fig.~\ref{fig:e_loc}(a) and (b) for 10 and 50 fs pulse width. The
result correlates very well with the energy distribution given in
Fig.~\ref{fig:Eatoms}, ~\ref{fig:Eelec}, and ~\ref{fig:density}. As
the energy of the particles and the density decreases, the temporary
localization increases. However, even in the worst case, approximately
50\% of the photons stay about 10~fs at a given site. Comparing this
to typical recombination times (1000~fs), the probability of the
recombination is small.

At last we would like to discuss shortly the consequences of our
calculations on the single molecule imaging suggested by R. Neutze et
al.\cite{Neutze} First, we would like to point out that the assumption
of neglecting the interaction of photo- and Auger electrons with the
cluster is justified only for small clusters ($<500~atoms$). As it is
clear from Fig.~\ref{fig:diffmodels}(c) the authors overestimated the
speed of the Coulomb explosion. This would mean an even higher
tolerance against radiation, and a better chance of successful imaging
by a single pulse. However, it is not enough to have the nuclei at
their proper positions, we must have electrons bound to the nuclei to
scatter x-ray photons. As we can see from
Fig.~\ref{fig:diffmodels}(a), the number of stripped electrons was
underestimated in Neutze's calculations. This error increases with
cluster size. To estimate the time available for imaging, we
calculated the deviation of temporary atomic configuration from the
original one (Fig.~\ref{fig:Rfact}). We took into account two
contributions: structural deviation (position changes, $\Delta r_i$)
and ionization state (how many electrons remain on the atoms to
scatter; changes in the atomic scattering factors $\Delta f_i$
). Allowing 20\% overall error we can measure up to $-40$ and $-6$ fs
in the case of 50 and 10~fs pulses respectively, to get useful
structural information. Since the pulse shape is gaussian, it is not
easy to visualize the meaning of these limits. A better
characterization can be given by the integral number of photons
incident on the sample within this period. We find that these are 3\%
and 10\% of the total number of photons in the case of a 50 and a
10~fs pulse, respectively. In practice this means that we have to
disable our detector during the major part of the pulse. A more
detailed analysis on the structural studies of small single particles
by hard x-ray free electron laser pulses will be given in a
forthcoming paper.

\section{\label{conclusion}Conclusions}

In this paper we gave a picture of the Coulomb explosion of a small
cluster of atoms initiated by a hard x-ray pulse. In the calculations
we covered a wide range of cluster size from 50 to 1500
particles/cluster for short (10~fs) and for long (50~fs) pulses. We
showed that the dynamics of the explosion is different from that of
the laser driven Coulomb explosion. The cause of this is twofold: (i)
ionization of atoms starts from the deepest core levels in contrast to
the laser case, where it starts from the weakly bound outer
shells. (ii) the high frequency of the electromagnetic field in the
x-ray case does not allow charged particles to gain appreciable
velocity along the field direction. These lead to the following
picture: Most of the electrons kicked out by the primary photoeffect
have high enough kinetic energy to leave the close environment of the
cluster well within the time width of the x-ray pulse. This results in
a positively charged cluster. However, this primary ionization is
enhanced by the Auger process and by the inelastic electron-atom and
electron-ion collisions. Electrons produced this way do not have
enough energy to leave the cluster immediately, a peculiar charge
distribution is created from highly charged ions and electrons. This
distribution is inhomogeneous; a closely neutral core is surrounded by
a positive shell. At the end of the pulse three typical energy
distributions can be distinguished: electrons in the inner almost
neutral core condense at low energies, most of the ions in this inner
part have also low energies, the remaining ions are in the positively
charged shell having a closely constant energy distribution at the
high energy side.

Beside the characterization of the Coulomb explosion, we also gave an
estimate for the useful time for structural imaging of small
clusters. We found that about the 3\% and 10\% of the total number of
photons in a pulse can be efficiently used for structural imaging from
a 50 and 10~fs pulse, respectively.

\begin{acknowledgments}
The work reported here was supported by the EU Centre of Excellence
program, OTKA T029931, T043237, and the EU BIO-4CT98-0377 grant. We
thank Janos Hajdu, Abraham Sz{\"o}ke, and Beata Ziaja for the illuminating
discussions.
\end{acknowledgments}


\begin{thebibliography}{32}
\bibitem[{osc()}]{osccalc}
\bibinfo{note}{In half period of the x-ray frequencies an electron changes its
  velocity by about $v=3*10^6~m/s$ (corresponding to $E_{kin}=30~eV$) and its
  position by $s=0.002~\textrm{\AA}$.}
\bibitem[{not()}]{notestartconf}
\bibinfo{note}{On precision we mean the systematic errors caused by the
  approximations (what processes are taken into account) used by the different
  models. Turning on and off a process we can get an estimate for the
  importance of that process.}
\bibitem[{loc()}]{locdef}
\bibinfo{note}{The number of the localized electrons was determined by going
  backwards in the calculations (so starting from the end), and counting those
  electrons, which stayed longer than 1,2,\ldots~etc fs in the vicinity of the
  same atom. Note that this way we can speak about localization only within the
  time frame of our calculation, we call this temporary localization.}
\end{thebibliography}

\end{document}